# Does Open Access Really Increase Impact? A Large-Scale Randomized Analysis

Abdelghani Maddi*, David Sapinho**

*abdelghani.maddi@hceres.fr
Observatoire des Sciences et Techniques, Hcéres, 2 Rue Albert Einstein, Paris, 75013 France

** david.sapinho@hceres.fr
Observatoire des Sciences et Techniques, Hcéres, 2 Rue Albert Einstein, Paris, 75013 France

**Introduction**

The Open Access Citation Advantage (OACA) has been a major topic of discussion in the literature over the past twenty years, generally assuming that a better accessibility fosters research impact. In fact, this belief is not so obvious, with contradictory findings in the studies dealing with the OACA, that have often proven tough to compare, depending on many factors such as disciplines, OA statuses taken into account, publications types as well as the method and database used.

Overall, studies that found the existence of OACA were more common in social sciences (Mikki *et al.*, 2018; Abbasi *et al.*, 2019; Valderrama-Zurián, Aguilar-Moya and Gorraiz, 2019), medical and health sciences (Hudson *et al.*, 2019; O'Kelly, Fernandez and Koyle, 2019; Miller *et al.*, 2021), and natural sciences (Lin, 2007; Wang *et al.*, 2015; Clements, 2017). OACA is less important (and in several disciplines non-existent) in Physical Sciences and Engineering (Archambault *et al.*, 2016). In addition, there are some studies (less numerous) that concluded on the nonexistence of OACA in social sciences (Zhang, 2006), medical and health sciences (Tonia *et al.*, 2016; Mimouni *et al.*, 2017; Nazim and Ashar, 2018), and natural sciences (Campos *et al.*, 2016; Peidu, 2020). On a large sample of the Web of Science database, (Dorta-González, González-Betancor and Dorta-González, 2017) concluded that there is no OACA in all disciplines.

Recently, a review analysis on the topic identified 134 publications dealing with OACA (Langham-Putrow, Bakker and Riegelman, 2021). Applying the EBL critical appraisal method to analyze the risks of bias based on factors like sample size, data collection or study design, the authors emphasized that most of these studies (131 – about 98%) present a high risk. Two of the three publications with a low risk of bias relate to medical and natural science research (Tonia *et al.*, 2016; Nelson and Eggett, 2017), and the third (Sotudeh, Ghasempour and Yaghtin, 2015) used relatively old data (2007-2011). Moreover, none of these articles used randomisation techniques/control group.

As several known confounding factors have limited the scope of the results, a key issue to address the question rigorously would be to figure out a way to "isolate" the OA effect. OA publications should thus be compared to a counterfactual sample of publications with the only difference is to be published in subscription-based journals.

In this paper, we propose a method to constitute a control group to isolate the OACA effect. By definition, OACA analysis answers the following question: "what would be the impact of an OA publication if it had been published in a subscription-based journal?" The answer this question is only possible if this publication is compared to its counterparts, with respect to characteristics that can influence the impact, such as the discipline, the type of journals, or other known characteristics (e.g. the number of authors, whether it has received a specific funding or the number of funders).

We limit ourselves to publications in gold OA, but distinguishing between publications in full OA journals and hybrid journals. The reason why we did not take into account the green and bronze routes is the difficulty of conducting an analysis of the OACA for this type of publications. Thus, for the green route, it is not possible to identify precisely the moment when each publication was dumped in an open archive when for the bronze route; the durability of the OA status is highly questionable.

**Data**

We extracted the publications data from the French OST in-house database. It includes five indexes of the WoS available from Clarivate Analytics (Science Citation Index Expanded (SCIE), Social Sciences Citation Index (SSCI), Arts & Humanities Citation Index (AHCI), Conference Proceedings Citation Index (CPCI-SSH) and Conference Proceedings Citation Index (CPCI-S)), and corresponds to WoS content indexed through the end of March 2021. The study focuses on three types of publications: articles, reviews and conference proceedings.

**Method**

*Sample selection procedure and weighting*
Figure 1 describes the sample selection procedure.

*OA sample*

We selected all the documents published with a Gold OA status from 2010 to 2020 by distinguishing those published in a full OA and those in hybrid journals, representing respectively 2,458,378 and 1,024,430 publications.

*Control sample*

We used the raking ratio method (Deville and Särndal, 1992; Deville, Sarndal and Sautory, 1993; Sebastian Daza, 2012) to ensure comparability between the two samples. The method comprises the construction of a counterfactual sample similar to the sample of interest, except for the analyzed parameter, which is in our case the OA status.

The control sample is thus obtained by finding all the non-OA publications, qualified as doubles, that match with OA publications on a set of features identified in the literature as having an effect on the citation impact (Judge *et al.*, 2007; Yan, Wu and Song, 2018; Waltman and Traag, 2021; Maddi and Sapinho, 2022). These features are characterised as follows:
- the publication year (11 classes : 2010 to 2020),
- the discipline (OST classification in 27 ERC panels),
- the journal impact (5 classes : <0.8, [0.8 , 1.2[, [1.2 , 1.8[, [1.8 , 2.2[, >=2.2), for the calculation method see: (Maddi and Sapinho, 2022).
- the number of countries of contributors, based on WoS addresses information (5 classes : 1,2,3,4 and 5 or more)
- the number of funding received, based on WoS acknowledgment information (5 classes : 1,2,3,4 and 5 or more)
- the presence of an ERC funding (2 classes : Yes or No)
- the presence of at least one European (UE27) address (2 classes : Yes or No)
- the presence of a patent citation (2 classes : Yes or No)

On this basis, we categorized each OA publication in one among 242,924 different clusters. From the same clusters, we then identify 12,088,681 double candidates among which 10,310,342 and 11,533,001 are respectively eligible for full OA and hybrid OA publications.

*Raking weights*

We finally calculated the raking weights for the publications of the control sample in order to ensure that both samples have a comparable distribution with respect to the characteristics identified.

In our calculations, we distinguish publications in full OA journals and those in hybrid journals.

Figure 1: Method of constructing the counterfactual sample of OA publications.

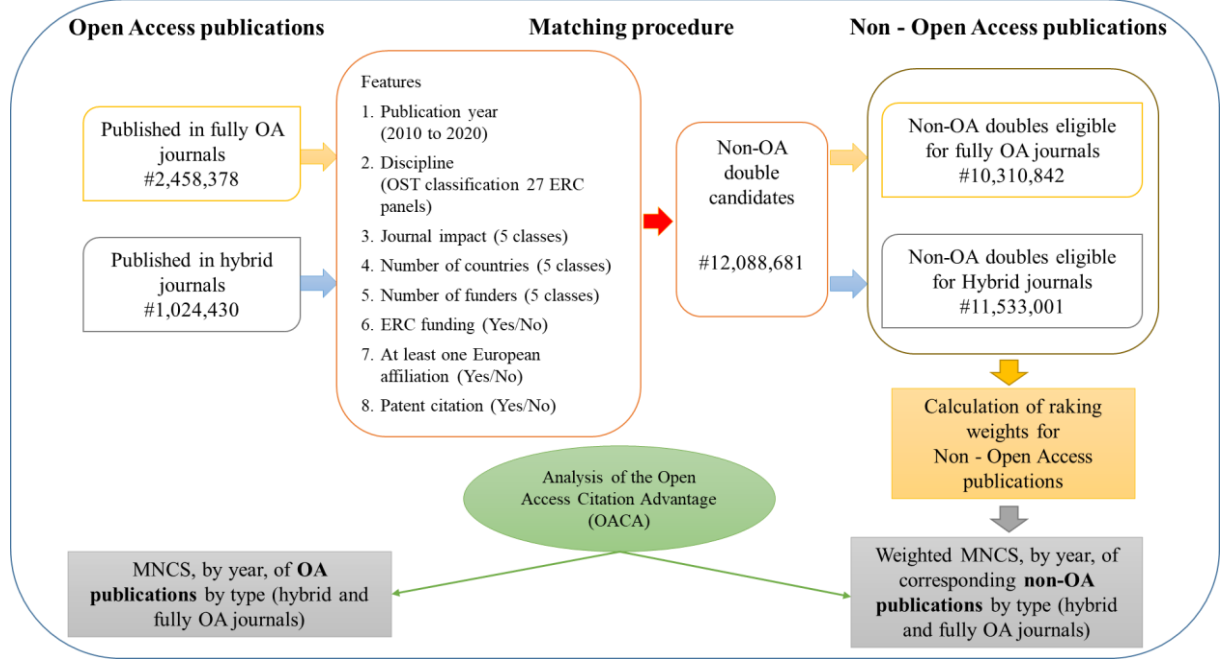

*OACA calculation*

We define the Open Access Citation Advantage (OACA) as the percentage of advantage in the citation score that could be attributed to Open Access (OA). We used the following formula:

$$OACA = \frac{MNCS_{OA} - MNCS_{Non-OA}}{MNCS_{Non-OA}} * 100$$

Where $MNCS_{OA}$ is the Mean Normalized Citation Score (Leydesdorff and Opthof, 2010) in the OA group, and $MNCS_{Non-OA}$, the same score in the Non-OA counterfactual group.

**Results**

Figure 2 describes the evolution of OACA of publications between 2010 and 2020, respectively in fully OA journals and hybrid journals. The first finding is that OACA highly depends on the type of journal. Indeed, for publications in fully OA journals, there is rather a citation disadvantage, which average about -20 to -15% over the period 2010-2020, when compared to the uncontrolled group (all non-OA WoS publications). This disadvantage reduced significantly until 2016 to approach zero before decreasing to reach the value at the start of the period (i.e. -20%).

Figure 2: OACA all disciplines combined

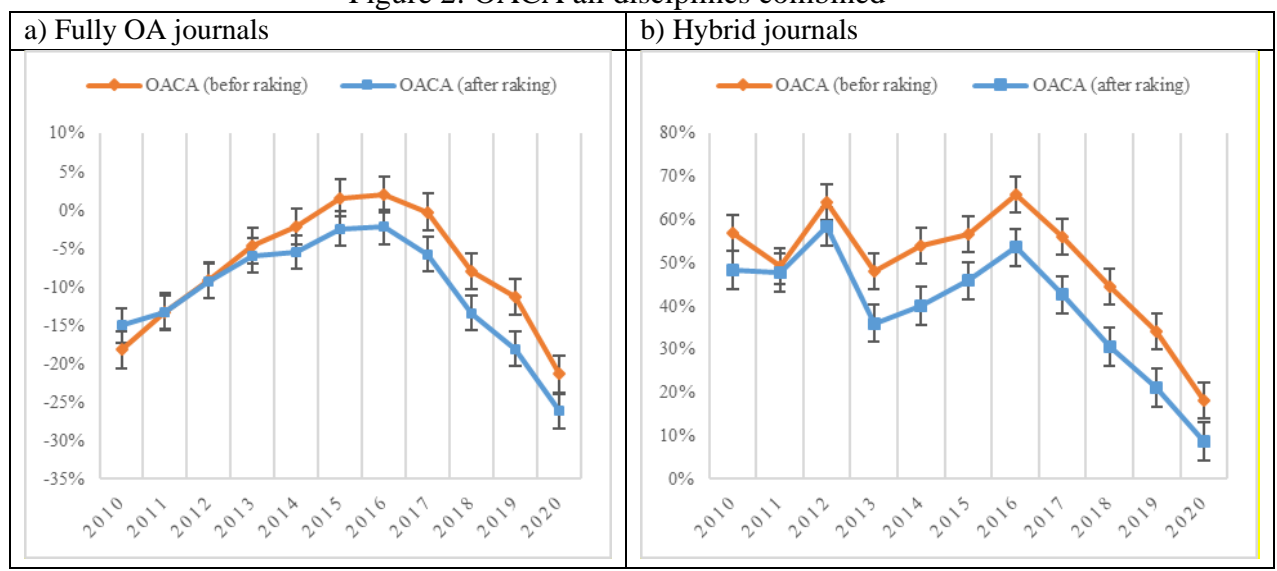

| a) Fully OA journals | b) Hybrid journals |

Contrasting this are the results for OA publications in hybrid journals. Figure 2 shows a high OACA at the start of the period, which receive, on average, 60% more citations than non-OA publications to reach 70% in 2016. Since that year, however, the downward trend is comparable to that observed for full OA journals. Even if OACA persists, it steadily declines throughout the end of the period.

This figure also highlights the effect of using a non-OA control group similar to that of OA publications represented by the blue curves. Whatever the type of journal, OACA tends to be lower with this group as reference, suggesting that the use of non-adjusted control groups overestimates OACA.

Figure 3 declines the OACA by discipline, comparing two periods, for each type of journal. In fully OA journals (Figure 3a), there is rather a citation disadvantage in most of disciplines, except for a few in Life Sciences. These are precisely "Immunity, Infection and Immunotherapy" (LS6), "Molecules of Life: Biological Mechanisms, Structures and Functions" (LS1), "Integrative Biology: From Genes and Genomes to Systems" (LS2) and "Cellular, Developmental and Regenerative Biology" (LS3). The OACA in these disciplines varies between 10 and 20% (after raking), in the second period (2018-20). The comparison between the two periods shows, in general, no major variations. Some exceptions should be noted, like PE2 for which the OACA dropped significantly from 50% in 2010-12 to 5% in 2018-20.

On the flipside, the OACA exists in the majority of disciplines in hybrid journals. LS6 is the discipline with the highest OACA (83% in the second period), followed by LS2 (73%). The OACA is however much lower (even negative in some cases) in Physical Sciences and Engineering. This is notably the case for Materials Engineering (PE11), Physical And Analytical Chemical Sciences (PE4) and Synthetic Chemistry And Materials (PE5) with OACA rates of -12, -13 and -14% respectively.

Figure 3: OACA (after raking) between 2010-12 and 2018-20, by discipline*

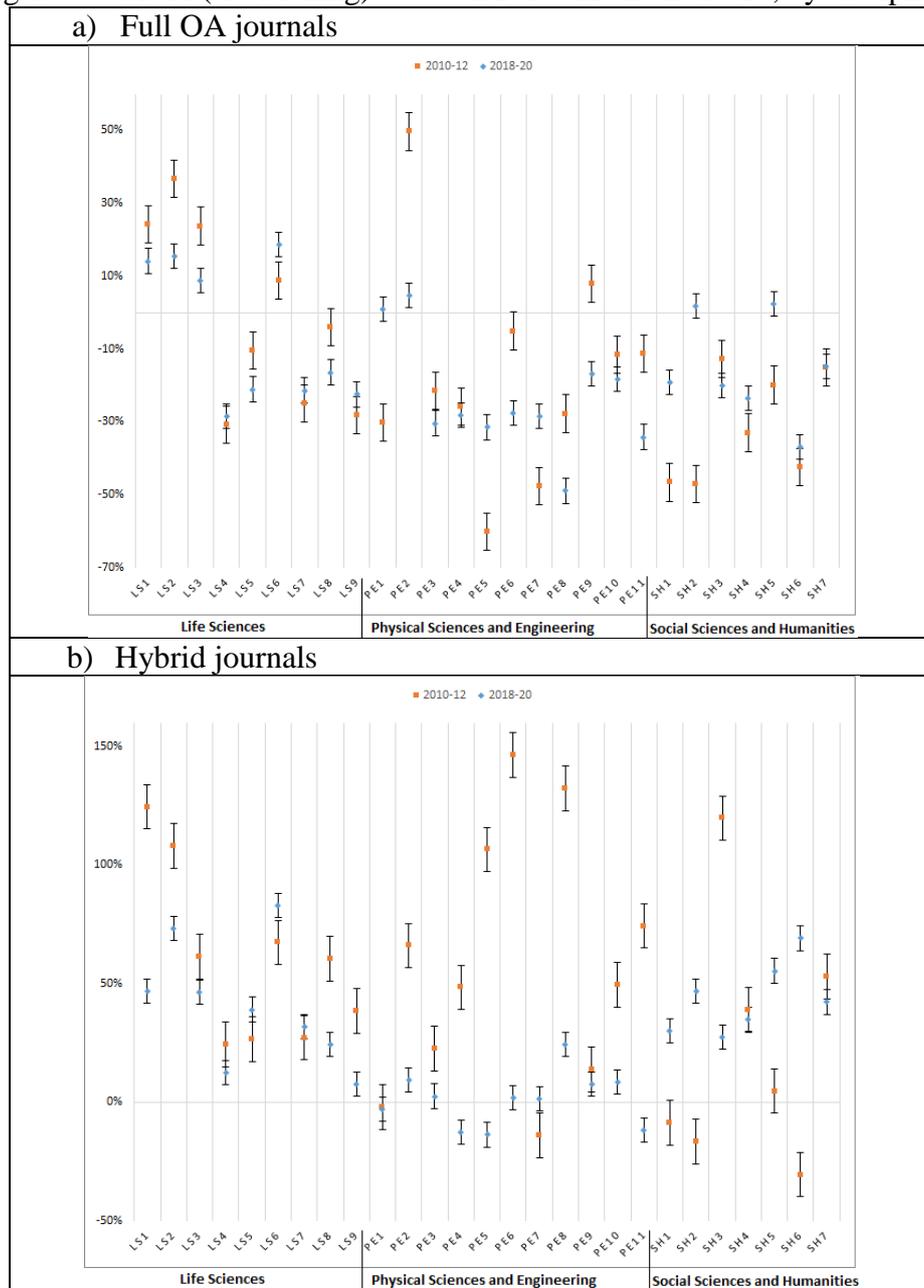

*For the labels of the disciplines see:
https://erc.europa.eu/sites/default/files/document/file/ERC_Panel_structure_2021_2022.pdf

The analysis of the evolution of the OACA over time makes it possible to deduce four main lessons:
- there are many disciplinary disparities concerning the evolution of the OACA over time,
- the greatest falls in the OACA are observed in the PE disciplines,
- on the contrary the OACA tends to increase in the SH disciplines,
- the OACA is relatively stable in the LS disciplines, except for LS1 and LS2 where a significant decrease is observed (and L8 and LS9 to a lesser extent).

**Conclusion**

Through this paper, we have analysed the question of Open Access Citation Advantage (OACA) in the Web of Science database. To do so, we compared citation impact (MNCS) of 2,458,378 publications in fully OA journals to that (weighted MNCS) of a control group of non-OA publications (#10,310,842). Similarly, we did the same exercise for OA publications in hybrid journals (#1,024,430) and their control group (#11,533,001), over the period 2010-2020. The results showed that there is no OACA for publications in fully OA journals, and that there is rather a disadvantage. Conversely, the OACA seems to be a reality in hybrid journals, suggesting that a better accessibility in this context tends to improve the visibility of publications.

Another important finding of our study is that 2016 appears to be a turning point on how an easier accessibility to publications influences their visibility, with a strong and regular downward trend from this year onwards, regardless of the type of journal.

Nevertheless, the analysis by discipline showed great discrepancies, both in the initial level of OACA and its evolution over the time. It could still be noticed that the most important falls are for disciplines where the initial level was the higher.

Finally, the effect of using an adjusted control group is clearly demonstrated, even if it has no effect in the trends observed.

**Discussion**

The lack of OACA for publications in fully OA journals is to be expected, as a great proportion of OA journals are newly created and less attractive to high-impact senior researchers. However, the citation disadvantage of these publications has largely diminished over time to disappear in 2016, as these journals gain a readership over the years and become more visible. OACA of publications in hybrid journals is also to be expected. These journals are well established in the publishing market and highly visible to scientific communities. For this type of journal, the fact that an article is in OA has a positive impact on citations more than articles that are in the same journals but with limited access to subscribers (hence the OACA). These results therefore suggest that the OACA depends on the type of journal; i.e. to hope to benefit from an OACA, it would first be necessary to choose a journal visible to the community (high-impact journal). In this case, the OA statue will act as a catalyst that would increase the citations received.

Another striking result of this paper is the fall of the OACA from 2016. The citation advantage fell from 70% to 9% between 2016 and 2020 (for publications in hybrid journals). We wonder if this fall is linked to the increase in the notoriety of pirate sites (eg Sci-Hub) from 2016 (especially after the ranking in 2016 of Alexandra Elbakyan - creator of Sci-Hub - in the top 10 influential figures of Nature (Nature's 10, 2016, p. 10)). In other words, the democratization of pirate sites instantly cancels the positive effect of OA publication insofar as the question of access to scientific content no longer arises. This is a hypothesis that we will investigate in future studies.